\documentclass[pdflatex,sn-mathphys-num]{sn-jnl}

\usepackage{graphicx}%
\usepackage{multirow}
\usepackage{array}
\usepackage{amsmath,amssymb,amsfonts}%
\usepackage{amsthm}%
\usepackage{mathrsfs}%
\usepackage[title]{appendix}%
\usepackage{xcolor}%
\usepackage{textcomp}%
\usepackage{manyfoot}%
\usepackage{booktabs}%
\usepackage{algorithm}%
\usepackage{algorithmicx}%
\usepackage{algpseudocode}%
\usepackage{listings}%
\usepackage{natbib}%

\theoremstyle{thmstyleone}%
\theoremstyle{thmstyletwo}%

\theoremstyle{thmstylethree}%

\begin{document}

\title[Article Title]{ScaleFusionNet: Transformer-Guided Multi-Scale Feature Fusion for Skin Lesion Segmentation}


\author*[1,2]{\fnm{Saqib} \sur{Qamar}}\email{sqama@kth.se}

\author[3]{\fnm{Syed} \sur{Furqan Qadri}}
\author[4]{\fnm{Roobaea} \sur{Alroobaea}}

\author[5]{\fnm{Goram Mufarah} \sur{M Alshmrani}}
\author*[5]{\fnm{Richard} \sur{Jiang}}\email{r.jiang2@lancaster.ac.uk}

\affil[1]{\orgdiv{Division of Robotics, Perception and Learning(RPL)}, \orgname{Department of Intelligent Systems}, \orgaddress{\street{KTH Royal Institute of Technology}, \postcode{10044}, \state{Stockholm}, \country{Sweden}}}

\affil[2]{\orgdiv{Faculty of Computing and Information Technology (FCIT)}, \orgname{Sohar University}, \orgaddress{\city{Sohar}, \postcode{311},\country{Oman}}}

\affil[3]{\orgdiv{BGI Research}, \orgaddress{\street{Hangzhou}, \postcode{310030}, \country{China}}}

\affil[4]{\orgdiv{Department of Computer Science}, \orgname{College of Computers and Information Technology}, \orgaddress{\street{Taif University}, \city{P. O. Box 11099}, \postcode{Taif 21944}, \country{Saudi Arabia}}}


\affil[5]{\orgdiv{School of Computing and Commutations}, \orgaddress{\street{ Lancaster University}, \city{Lancaster LA1 4YW}, \country{UK}}}

\abstract{Melanoma is a malignant tumor that originates from skin cell lesions. Accurate and efficient segmentation of skin lesions is essential for quantitative analysis but remains a challenge owing to blurred lesion boundaries, gradual color changes, and irregular shapes. To address this, we propose ScaleFusionNet, a hybrid model that integrates a Cross-Attention Transformer Module (CATM) and adaptive fusion block (AFB) to enhance feature extraction and fusion by capturing both local and global features. We introduce CATM, which utilizes Swin transformer blocks and Cross Attention Fusion (CAF) to adaptively refine feature fusion and reduce semantic gaps in the encoder-decoder to improve segmentation accuracy. Additionally, the AFB uses Swin Transformer-based attention and deformable convolution-based adaptive feature extraction to help the model gather local and global contextual information through parallel pathways. This enhancement refines the lesion boundaries and preserves fine-grained details. ScaleFusionNet achieves Dice scores of 92.94\% and 91.80\% on the ISIC-2016 and ISIC-2018 datasets, respectively, demonstrating its effectiveness in skin lesion analysis. Simultaneously, independent validation experiments were conducted on the PH$^2$ dataset using the pretrained model weights. The results show that ScaleFusionNet demonstrates significant performance improvements compared with other state-of-the-art methods. Our code implementation is publicly available at \href{https://github.com/sqbqamar/ScaleFusionNet}{GitHub}.}




\maketitle

\section{Introduction}\label{sec1}
 The incidence of melanoma has risen significantly in recent decades due to increasing environmental pollution and ultraviolet radiation \cite{storelvmo_assessing_2025}. This trend has attracted significant attention from the global medical community. Early detection is essential for improving treatment outcomes and patient survival rates. Traditional diagnostic methods, such as clinical observation and tissue biopsy, are limited by subjectivity and invasiveness, making them unsuitable for large-scale screening. Medical image segmentation offers a noninvasive and high-precision alternative that provides clinicians with more detailed and accurate information. This technology has the potential to significantly enhance the early detection and treatment of skin cancer. 

With the rapid advancement of deep learning, convolutional neural networks (CNNs) have achieved notable success in medical image segmentation tasks such as skin lesion segmentation. Among these, the U-Net model \cite{navab_u-net_2015} stands out as a pioneering framework. U-Net has demonstrated remarkable performance in medical image segmentation and established a U-shaped architectural paradigm. However, CNN-based methods often struggle to capture global contextual information effectively due to the inherent limitations of convolutional operations. This limitation is particularly evident in medical image segmentation tasks with significant inter-sample variability, such as skin lesion segmentation. To address this challenge, researchers have explored various strategies, including the use of large kernel convolutions, dilated convolutions, and other techniques aimed at expanding receptive fields \cite{cui_p2tc_2025, singh_less_nodate, li_single_2024, hu_lamffnet_2025, qamar2021dense}. For example, Hu et al. \cite{hu_mlda-net_2025}  improved receptive fields by using self-attention, while Tang et al. \cite{tang_cmunext_2024} proposed a model that used large convolutional kernels and fusion to achieve promising results in tasks such as breast nodule ultrasound image segmentation. Inspired by the ConvNeXt model \cite{liu_convnet_2022}, Han et al. \cite{han_convunext_2022} developed a method for medical image segmentation using large kernel convolutions and they successfully implemented it for tasks such as retinal vessel segmentation. Despite these advancements, simply increasing the size of the convolutional kernels may not fully resolve the challenge of modelling global features because the fundamental constraints of the receptive field remain. 

Recently, Transformers \cite{thirunavukarasu_comprehensive_2024} have achieved notable success in both the natural language processing and computer vision domains by using global contextual information in feature extraction. Cai et al. \cite{cai_intelligent_2024} unveiled the BiADATU‐Net that combined Transformer and feature adaptation modules, which resulted in promising outcomes in several publicly accessible skin lesion segmentation datasets. Zhang et al. \cite{zhang_dual-attention_2024} developed DAE-Former, a pure Transformer U-shaped medical image segmentation model, harnessing efficient Transformers steered by dual attention, similar to Swin-Unet \cite{karlinsky_swin-unet_2023}, and exhibited commendable performance in diverse image segmentation datasets, including ISIC-2018 \cite{codella_skin_2019}. However, because Vision Transformers can only output single-scale feature representations, they lack the ability to capture multi-scale information in two-dimensional images \cite{broedermann_hrfuser_2023, wang_xbound-former_2023}. Consequently, transformer-based medical image segmentation models may struggle to seamlessly integrate multi-scale information, leading to insufficient attention to lesion regions and incomplete decoding of feature details. Additionally, there is a significant issue with medical image segmentation models based on the U-Net design architecture. Although skip connections in U-Net transmit multi-scale information between different stages to the decoder, a semantic gap issue may arise when there is a considerable semantic difference between encoder and decoder. To address this, some studies have attempted to mitigate this issue by improving skip connections. For instance, UNet++ \cite{stoyanov_unet_2018} and MISSFormer \cite{huang_missformer_2021} aim to achieve the fusion of multi-scale information between different stages through dense skip connections and contextual bridges. Nevertheless, this study argues that the differently sized feature maps transmitted through skip connections represent macroscopic multi-scale information that is easily observable, and such methods have limited effectiveness in enhancing the model’s ability to integrate multi-scale information. In particular, in the skin lesion segmentation task, the lesion edges are often irregular, with colors gradually fading from the center, and the progressive compression of feature maps leads to the loss of fine details, retaining only a macro-level focus. This can affect the model performance to some extent.

To address the challenges of skin lesion segmentation, this study proposes ScaleFusionNet, a model that integrates an AFB and CATM for enhanced feature extraction and fusion. ScaleFusionNet employs a hierarchical Swin Transformer-based encoder, where patch embedding and Swin Transformer blocks \cite{islam_cost-unet_2024} extract the multi-scale features. The decoder utilizes AFBs, which combine Swin Transformer and deformable convolution features to refine feature integration and improve lesion boundary preservation. To bridge the semantic gap between the encoder and decoder features, the CATM uses cross-attention, which allows high-level decoder features to guide low-level skip connections. The experimental results demonstrate that ScaleFusionNet achieves highly competitive performance in skin lesion segmentation. The key contributions of this study are as follows:

\begin{itemize}
    \item We have proposed ScaleFusionNet for skin lesion segmentation, based on a hybrid architecture combining CNNs and Transformers, which outperforms other state-of-the-art methods.
    \item We have introduced AFB, which integrates both Swin Transformer-based and deformable convolution-based feature extraction, enabling the model to capture both local and global contextual information.
    \item We developed the CATM to effectively reduce the semantic gap and enhance the interaction between the encoder and the decoder.
\end{itemize}

\section{Related Work}\label{sec2}

\subsection{CNNs for Medical Image Segmentation}
Recently, CNNs have achieved success in different domains due to their powerful feature extraction capabilities. This success is particularly evident in medical image segmentation. In 2015, Ronneberger et al. introduced U-Net, a CNN-based model designed specifically for medical image segmentation, which has become foundational in this field. Zhou et al. \cite{unet++} proposed UNet++, which introduces nested dense skip connections to address the semantic gap between the encoder and decoder. Li et al. \cite{li_h-denseunet_2018} developed an H-DenseUNet, a U-shaped model that enhances intra-slice and inter-slice representations through hybrid dense connections, demonstrating effective performance in liver tumor segmentation tasks. Saqib et al. \cite{qamar2021hi} proposed multi-scaled architecture using separable convolution for brain tumor segmentation. Oktay et al. \cite{oktay_attention_2018} developed Attention U-Net, a model that focuses on important areas by using attention gates, which helps reduce the differences between the encoder and decoder. Furthermore, UNet3+ \cite{huang_unet_2020} advanced the skip connection by adding full-scale skip connections and deep supervision, achieving better results in segmentation tasks. Xie et al. \cite{xie2024} designed a feature-steered network to learn the more distinctive features, which is built on a scale-adaptive module and cross path fusion (CPF) module. Wang et al. \cite{wang2025skin} used attention-based UNet to enhance the completeness of representation with the fusion of edge and body features. Katar et al. \cite{katar2025att} introduced a mixed model that combines ConvNeXt blocks with self-attention methods to improve skin lesion segmentation. Despite the success of U-Net and its variants, CNN-based models are limited by their inability to capture long-range dependencies. This limitation arises from the inherent nature of convolution operations, which struggle to model global contextual information. Additionally, dense skip connections based on simple summation offer limited solutions for addressing the semantic gap, particularly in tasks where fine-grained detail and global context are crucial, such as skin lesion segmentation.

\subsection{Transformers for Medical Image Segmentation}
Transformers, which are adept at capturing long-range dependencies, offer an effective alternative to CNNs. The Vision Transformer \cite{fan_multiscale_2021}, the first application of Transformer to computer vision, partitions input images into a sequence of patches for embedding and encoding using Transformer blocks. This innovative approach inspired Transformer-based U-shaped models for medical image segmentation tasks. TransUnet (\cite{chen_transunet_2021}) integrates transformer blocks with U-Net, leveraging their global feature modelling capabilities. In parallel, Swin-Unet \cite{karlinsky_swin-unet_2023} drew on the Swin Transformer to propose a fully Transformer-based method that applies Swin Transformer blocks to both the encoder and decoder. For 3D medical image segmentation, nnFormer \cite{zhou_nnformer_2022} combines local and global attention for multi-organ segmentation, demonstrating impressive performance. However, transformer models like TransUnet have problems because they have a lot of parameters and are complicated to compute, and the Swin transformer’s shifting window method can create rough edges in some situations. Furthermore, transformers inherently operate on single-scale outputs, limiting their ability to fully utilize multi-scale information, which hampers performance, especially in segmentation tasks requiring precise boundary delineation.

\subsection{Deformable Convolution Network}
The Deformable Convolutional Network (DCN) \cite{dai_deformable_2017} is an extension of traditional convolutional networks designed to improve the adaptability of convolutional kernels to object deformations. DCNs introduce learned offsets to control the sampling positions of kernels on input feature maps. This dynamic adjustment allows the model to better capture the features of targets with varying shapes and positions. Compared to traditional convolution, deformable convolutions offer more flexibility in capturing features of objects with diverse shapes and scales, which is crucial in medical image segmentation. DCNs have proven effective in addressing the complexities of various segmentation tasks, where object deformations are a significant concern \cite{mu_deformable_2025}. Xin et al. \cite{yang2022dcu} used deformable convolution to build a feature extraction module, which enhances the modeling ability of the model for deformation. However, their method shows limited generalization on complex structures. 

\begin{figure}[h]
    \centering
    \includegraphics[width=0.8\textwidth]{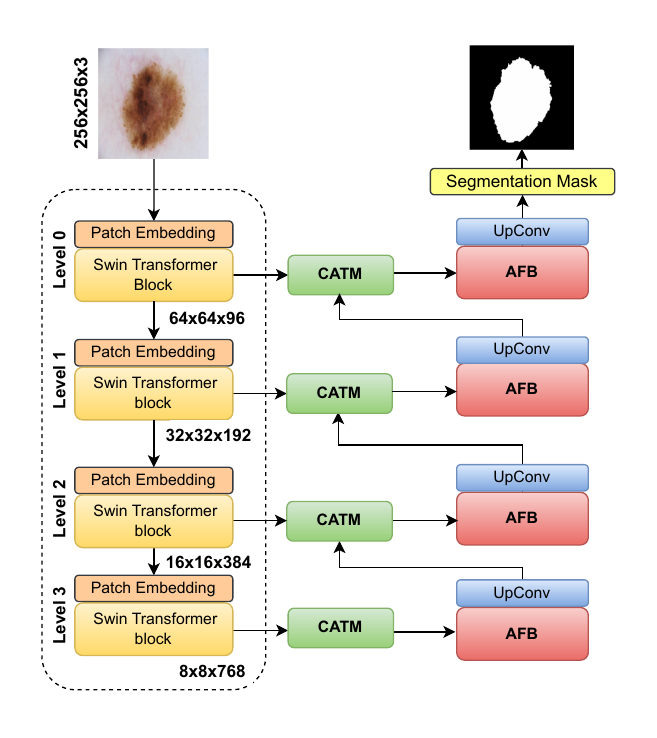}
    \caption{Architecture of ScaleFusionNet with a U-Net design, consisting of an encoder, CATM, and AFB. The encoder, utilizing convolutional layers and Swin transformer blocks, extracts multi-scale features at resolutions $64×64×96$, $32×32×192$, $16×16×384$, and $8×8×768$. The CATM refines feature fusion at skip connections using cross-attention, while the AFB enhances multi-scale fusion with deformable convolutions and Swin transformer-based attention to preserve fine-grained details for accurate segmentation.}
    \label{fig1}
\end{figure}

Recently, Ma et al. \cite{ma2024u} presented U-Mamba for general-purpose biomedical image segmentation, which integrates the advantages of local pattern recognition from CNNs and global context understanding from Mamba. Compared to CNNs, U-Mamba's SSM-based latent states are less interpretable than layer-wise CNN feature visualizations, and unlike Transformers, it lacks explicit attention maps for global context analysis, complicating debugging and trust in clinical settings. Li et al. \cite{li2024transiam} presented a dual-path network that uses two parallel CNNs for feature extraction from different modalities. It integrates CNNs and Transformers with a feature-level fusion strategy that uses the local attention aggregation block to focus on region-of-interest features and suppress invalid regions. However, it does not address the semantic gap between features. To achieve better accuracy in medical image segmentation, especially for skin lesions, ScaleFusionNet combines multi-scales of feature extraction with a hybrid of Transformer and CNN designs to improve how well it segments images. Traditional CNN-based models like U-Net struggle with fine-grained details and global context, prompting the need for improved architectures. ScaleFusionNet addresses these issues by using AFBs that incorporate swin transformers and deformable convolutions to collect features at various scales, which helps to refine the edges of lesions while preserving small details. Additionally, the CATM enhances encoder-decoder feature fusion, reducing semantic gaps through guided attention mechanisms. By combining Swin transformer blocks \cite{islam_cost-unet_2024} for global context and adaptive multi-scale fusion for local detail refinement, ScaleFusionNet achieves superior segmentation accuracy, demonstrating strong generalization in the skin lesion segmentation task.

\section{Methods}\label{sec3}

Figure~\ref{fig1} illustrates the architecture of ScaleFusionNet, which follows the U-Net design and consists of three primary components: an encoder, CATM, and AFB. The encoder employs a hybrid approach that integrates convolutional layers and Swin transformer blocks to effectively capture local and global features. By combining convolutional locality with self-attention mechanisms, the encoder enhances feature representation, ensuring robust extraction of hierarchical information. The CATM is introduced at skip connections to refine encoder-decoder feature fusion. It utilizes Swin transformer blocks and a CAF to mitigate the semantic gap and dynamically align hierarchical features. By using cross-attention mechanisms, the CATM enhances the integration of skip connection features with decoder information, ensuring improved contextual understanding. The AdaptiveFusionBlock further enhances multi-scale feature extraction and fusion by integrating deformable convolutions with Swin transformer-based attention. This fusion process refines the lesion boundaries and preserves fine-grained details, which are essential for accurate segmentation. Given an input image \( I \in \mathbb{R}^{H \times W \times C} \), where \( H \), \( W \), and \( C \) denote the height, width, and number of channels, respectively, the encoder progressively extracts the multi-scale features. The CATM is positioned at skip connection points, receiving features \( X_{\text{Skip}} \) from the encoder and \( X_{\text{Decoder}} \) from a lower-level decoder. The extracted features are dynamically refined using cross-attention mechanisms, allowing for precise alignment and feature enhancement. The AFB processes these refined features by fusing multi-scale information through deformable convolutions and Swin transformer-based attentions. The final segmentation result is obtained after feature fusion and upsampling operations in the decoder, ensuring a high-resolution and well-defined segmentation mask.

\begin{figure}[h]
    \centering
    \includegraphics[width=0.8\textwidth]{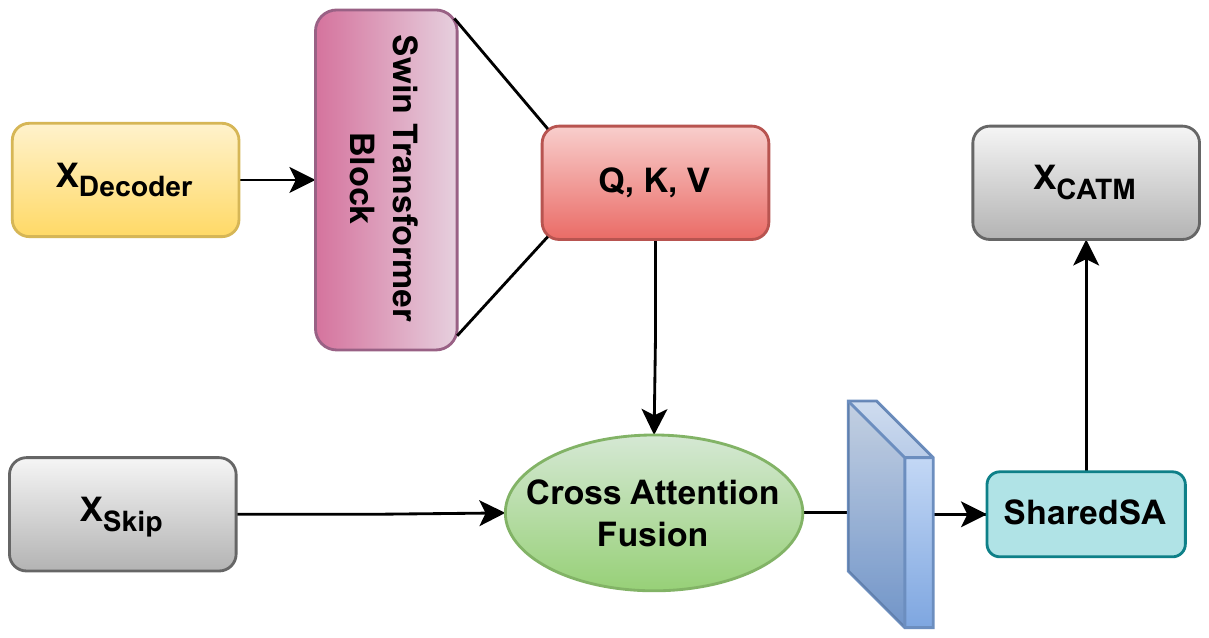}
    \caption{Schematic diagram of CATM. The decoder features $X_{\text{Decoder}}$ generate query, key, and value representations, which are fused with the encoder features $X_{\text{Skip}}$ via CAF. The refined features are then processed with SharedSA to produce the final aligned features $X_{\text{CATM}}$, preserving both fine details and high-level semantics.}
    \label{fig2}
\end{figure}

\subsection{CATM}\label{subsec1}

In U-Net, skip connections serve to provide information supplementation. During encoding, continuous compression of feature maps leads to a significant loss of spatial detail. Using low-level semantic features from the encoder to supplement the decoder feature restoration is an effective strategy. However, a fundamental issue remains: the semantic gap between the encoder and decoder features. Only concatenating features at different semantic levels can result in performance degradation owing to this misalignment.

To address this, we introduced the CATM to refine the encoder-decoder feature fusion. Unlike conventional skip connections, the CATM employs Swin transformer blocks and cross-attention fusion to adaptively align hierarchical features. By using self-attention and cross-attention mechanisms, the CATM ensures the effective transfer of relevant spatial and contextual information across the network. As shown in Figure~\ref{fig2}, given the encoder features \( X_{\text{Skip}} \) and decoder features \( X_{\text{Decoder}} \), the CATM dynamically refines the feature alignment using a learnable attention mechanism. This process is formulated as follows.

\[
Q, K, V = \text{Swin Transformer Block}(X_{\text{Decoder}})
\]

\[
X'_{\text{Skip}} = \text{CAF}(X_{\text{Skip}}, Q, K, V)
\]
Here, the Swin transformer block extracts query, key, and value representations from the decoder features, whereas the CAF integrates these with the encoder features\( X_{\text{Skip}} \), thereby reducing the semantic gap. After passing through the CAF, the features \( X'_{\text{Skip}} \) undergo a full-stage parameter-shared spatial attention mechanism to achieve unified feature attention, ultimately yielding features that are guided by high-level semantics.

\[
X_{\text{CATM}}= \text{SharedSA}(X'_{\text{Skip}})
\]
Where SharedSA represents the spatial attention operation shared across all stages, and \( X'_{\text{Skip}} \) denotes the result after passing through the CAF. The refined features \( X_{\text{CATM}} \) preserve both fine-grained details and high-level contextual information.

\begin{algorithm}[h]
\caption{CATM for ScaleFusionNet}
\label{alg:catm}
\begin{algorithmic}[1]
\Require \( X_{\text{Decoder}}, X_{\text{Skip}} \in \mathbb{R}^{H \times W \times C} \)
\Ensure \( X_{\text{CATM}} \in \mathbb{R}^{H \times W \times C} \)

\State \( Q, K, V = \text{SwinTransformerBlock}(X_{\text{Decoder}}) \)  \Comment{Obtain Query, Key, and Value}
 
\State \( X'_{\text{Skip}} = \text{CAF}(X_{\text{Skip}}, Q, K, V) \) \Comment{Feature Alignment}

\State \( X_{\text{CATM}} = \text{SharedSA}(X'_{\text{Skip}}) \) \Comment{Refined Feature Fusion}

\State \Return \( X_{\text{CATM}} \) \Comment{Return}
\end{algorithmic}
\end{algorithm}

\subsection{AFB}\label{subsec3}

The decoder plays a critical role in feature decompression and mask generation for medical image segmentation. A key challenge is the accurate restoration of boundary details and enhancement of attention toward target regions. Conventional decoder designs, whether convolution- or transformer-based, often struggle to effectively capture fine-scale information, leading to imprecise lesion localization and segmentation. To address these limitations, we introduced the AFB, which integrates adaptive multi-scale feature fusion to refine segmentation. This block is built with Swin transformer-based attention and deformable convolution-based adaptive feature extraction, allowing the model to capture both local and global contextual information through parallel pathways.

\begin{figure}[h]
    \centering
    \includegraphics[width=0.8\textwidth]{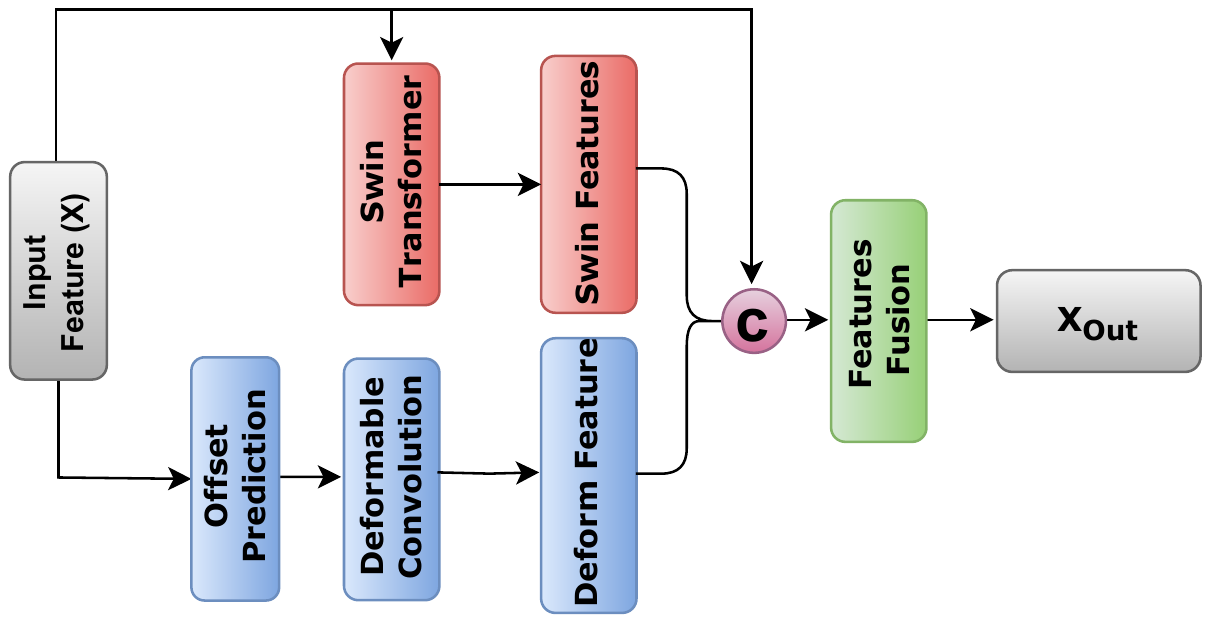}
    \caption{Schematic diagram of AFB for multi-scale feature fusion in the decoder. The input feature X is processed through three parallel branches: the Swin transformer branch $X'_{\text{Swin}}$, deformable convolution branch $X'_{\text{Deform}}$, and identity branch. The outputs are concatenated $X_{\text{Combined}}$ and passed through a $1$x$1$ convolution to produce the final output $X_{\text{Out}}$, refining lesion boundaries and improving segmentation accuracy.}
    \label{fig3}
\end{figure}

As shown in Figure~\ref{fig3}, given the input feature ($X$) from the decoder, the AFB processes it through three parallel branches to extract complementary representations: Swin transformer, deformable convolution, and identity branches. In the Swin transformer branch, the model employs a resolution-aware adaptation strategy to balance efficiency and richness of features. The Tiny variant of Swin transformer is used to capture long-range dependencies, but its processing is dynamically adjusted based on the encoder level and spatial resolution. The Swin transformer branch employs a resolution-aware adaptation strategy to accommodate varying spatial resolutions during the decoding process. Specifically, for higher-resolution inputs (Level 0: 64x64), only the first two Swin Transformer stages are utilized to preserve fine-grained, detailed features. At intermediate resolutions (Level 1: 32x32), the model processes the features through the first three Swin stages, providing a balance between the feature richness and computational cost. For deeper levels with lower resolutions (Level 2: 16x16), all four Swin Transformer stages are employed; however, the embedding dimensions are reduced to mitigate potential memory bottlenecks. At the deepest level (Level 3: 8x8), the Swin Transformer is omitted altogether, and a simple convolutional operation is applied instead because the spatial resolution at this stage is too limited to benefit from self-attention mechanisms. This adaptive approach ensures computational efficiency while preserving the benefits of hierarchical feature learning.

\[
X'_{\text{swin}} =
\begin{cases}
\text{SwinTransformer}(X), & \text{if } \text{encoder\_level} < 3 \\
\text{Conv}(X), & \text{otherwise}
\end{cases}
\]

The deformable convolution branch performs spatially adaptive feature extraction with offset prediction as follows:
\[
\text{Offset} = \text{Conv2D}_{(3 \times 3)}(X)
\]
\[
X'{\text{Deform}} = \text{DeformConv}(X, \text{Offset})
\]
The identity branch preserves the original features X to maintain low-level information for the model stability. The outputs from all three branches are fused through channel-wise concatenation, followed by feature reduction:
\[
X_{\text{Combined}} = \text{Concat}(X, X'{\text{Swin}}, X'{\text{Deform}})
\]
\[
X_{\text{Out}} = \text{Conv2D}_{(1 \times 1)}(X{\text{Combined}})
\]
This three-way fusion mechanism enhances the refinement of the lesion boundary while preserving fine-grained details through complementary feature representations. The deformable convolution adapts to irregular lesion shapes through learnable spatial offsets, the Swin transformer provides a global contextual understanding, and identity mapping maintains essential low-level features.
By integrating AFB into ScaleFusionNet, we achieved improved segmentation performance by effectively combining multiple feature extraction strategies in parallel, leading to better lesion delineation and generalization across datasets.

\begin{algorithm}[h]
\caption{AFB for ScaleFusionNet }
\label{alg:adaptivefusion_corrected}
\begin{algorithmic}[1]
\Require \( X \in \mathbb{R}^{H \times W \times C} \)
\Ensure \( X_{\text{Out}} \in \mathbb{R}^{H \times W \times C} \)

\State \( X'_{\text{Swin}} = \text{SwinTransformer}(X) \) \Comment{Extract features using Swin Transformer}

\State \( \text{Offset} = \text{Conv2D}_{3 \times 3}(X) \) \Comment{Predict offset for deformable convolution}

\State \( X'_{\text{Deform}} = \text{DeformConv}(X, \text{Offset}) \) \Comment{Extract features using deformable convolution}

\State \( X_{\text{Combined}} = \text{Concat}(X, X'_{\text{Swin}}, X'_{\text{Deform}}) \) \Comment{Concatenate original and extracted features}

\State \( X_{\text{Out}} = \text{Conv2D}_{1 \times 1}(X_{\text{Combined}}) \) \Comment{Fuse features using 1x1 convolution}

\State \Return \( X_{\text{Out}} \) \Comment{Return the fused output features}
\end{algorithmic}
\end{algorithm}

\section{Experiments}\label{sec4}
\subsection{Datasets}\label{subse4}
\textbf{ISIC-2016:} This dataset is derived from the skin lesion analysis for the melanoma detection challenge in 2016, comprising 1250 images meticulously annotated by professional experts with high-quality standard labels \cite{gutman_skin_2016}. Among these, 900 images are designated as training data, and 350 images are allocated for validation.

\noindent\textbf{ISIC-2018:} The ISIC-2018 dataset, also collected by ISIC in 2018, consists of 2594 images and corresponding labels. The resolutions of the images ranged from \(720 \times 540\) to \(6708 \times 4439\) pixels \cite{codella_skin_2019}. Among these, 2594 images are randomly divided into training, validation, and test sets at a ratio of 8:1:1.

\noindent\textbf{PH$^2$:} The dermoscopic images used in this study were acquired from Hospital Pedro Hispano in Matosinhos, Portugal, using the Tuebinger Mole Analyzer system set at 20x magnification. These images are formatted as 8-bit RGB color files with dimensions of 768×560 pixels. The dataset consists of 200 dermoscopic images featuring various melanocytic lesions.

\subsection{Evaluation metrics}\label{subse4a}
The main evaluation metrics used the Dice coefficient (DSC), Intersection over Union (IOU), Sensitivity (SE), Specificity (SP), and Accuracy (ACC).
The Dice coefficient measures the overlap between the predicted segmentation mask and the ground truth mask. It is defined as:
\[
DSC = \frac{2 |A \cap B|}{|A| + |B|}
\]
where \( A \) is the predicted mask and \( B \) is the ground truth mask. A higher Dice score indicates better segmentation performance.
Intersection over Union evaluates segmentation accuracy by computing the ratio of the intersection to the union between the predicted and actual masks:
\[
IOU = \frac{|A \cap B|}{|A \cup B|}
\]
SE measures the proportion of actual positive samples correctly identified by the model. 
\[
\text{SE} = \frac{|A \cap B|}{|B|}
\]
SP measures the proportion of actual negative samples that the model correctly identifies.
\[
\text{SP} = \frac{|\overline{A} \cap \overline{B}|}{|\overline{B}|}
\]
where $\overline{A}$ and $\overline{B}$ are compelments of A and B. Universal set U is the total number of pixels in the image.
ACC measures the proportion of correct predictions made by a model out of total predictions.
\[
\text{ACC} = \frac{|A \cap B| + |\overline{A} \cap \overline{B}|}{|U|}
\]

\subsection{Implementation details}
All experiments in this paper were conducted using the PyTorch 1.12.0 framework. The experiments were performed on a computer equipped with an Ubuntu 18.04 operating system, Intel Core i9-13900K CPU, Nvidia RTX 4060 GPU, and 1TB solid-state drive. For all experiments involving ScaleFusionNet, the AdamW optimizer was utilized with a learning rate and weight decay set to 1e-4. We used a combination of BCE and IOU losses to form our loss function, along with random rotation and random flipping for data augmentation. For comparison, we referenced the experimental results disclosed in relevant papers for similar methods. For outstanding models that did not perform skin lesion segmentation tasks, we retrained them using publicly available official execution codes. To ensure fairness, we kept parameters that did not affect the model learning capacity, such as epochs and batch size, consistent with ScaleFusionNet, setting epochs to 200 and batch size to 8. The input size for the network was 256 × 256. The experiments focused on the ISIC-2016 and ISIC-2018 datasets,and external testing was conducted on the PH$^2$ dataset based on the trained weights.

This experimental setup ensured a robust and fair evaluation of ScaleFusionNet's performance by utilizing state-of-the-art hardware and software configurations to achieve accurate and reproducible results. The use of a combined loss function and data augmentation techniques further enhances the model's ability to generalize and perform well on diverse skin lesion segmentation tasks.

\section{Experimental Results}
\subsection{Results on ISIC-2016}

We selected 13 prominent models for comparison with the proposed ScaleFusionNet. In all 13 models, we also included SAM2-UNet \cite{xiong2024sam2} and U-Mamba \cite{ma2024u} for skin lesion segmentation. The SAM2-UNet is an emerging vision foundation model that continuously achieves good performance on various tasks. UMamba is a general-purpose network inspired by State Space Sequence Models (SSMs), a new family of deep sequence models known for their strong capability in handling long sequences. Table \ref{tab:isic2016} shows that ScaleFusionNet achieved a DSC score of 92.94\% and an IOU score of 87.35\% on the ISIC-2016 dataset, which are the average results of five-fold experiments, demonstrating outstanding performance. Compared with the Swin-Unet model, ScaleFusionNet improved by 2.82\% in the DSC metric and 4.14\% in the IOU metric. Compared with MISSFormer, ScaleFusionNet exhibited enhancements of 2.48\% and 3.43\% in the DSC and IOU metrics, respectively. Against the D-LKA model, ScaleFusionNet still showed improvements of 0.11\% and 0.20\% in the DSC and IOU metrics, respectively. This indicates that ScaleFusionNet is more accurate than D-LKA in detecting the refined boundary of skin lesions.  Although the performance gains may not be significant, from the perspective of parameter count and computational complexity, ScaleFusionNet reduces the number of parameters by 33.2\% and decreases the computational load by 17.5\% compared with D-LKA. This indicates that ScaleFusionNet consumes far fewer hardware resources than D-LKA while maintaining the model size and computational complexity. SAM2Unet and U-Mamba are also behind ScaleFusionNet in terms of DSC and IOU. Compared to other methods such as U-Net, although ScaleFusionNet employs a more complex architecture to address potential issues in U-shaped medical image segmentation models, we find this approach justified given the 5.13\% performance improvement and reduced computational load. In clinical applications, a faster and lighter model can support a broader range of compatible use cases, which is crucial for hospitals and organizations with limited computational resources.

\begin{table}[h]
\centering
\caption{Performance Comparison on ISIC-2016 Dataset}\label{tab:isic2016}
\renewcommand{\arraystretch}{1.5}
\begin{tabular}{lcccc}
\toprule
\textbf{Methods} & \textbf{Params(M)} & \textbf{FLOPs(G)} & \textbf{DSC} & \textbf{IOU} \\
\midrule
U-Net \cite{navab_u-net_2015} & 34.53 & 124 & 87.81$\pm$0.41 & 80.25$\pm$0.50 \\
Att-Unet \cite{oktay_attention_2018} & 34.88 & 126.1 & 87.43$\pm$0.47 & 79.70$\pm$0.62 \\
nnU-Net \cite{isensee_nnu-net_2021} & - & - & 90.45$\pm$0.35 & 84.52$\pm$0.53 \\
SwinUNet \cite{karlinsky_swin-unet_2023} & 27.17 & 6.16 & 90.12$\pm$0.39 & 83.21$\pm$0.57 \\
MISSFormer \cite{huang_missformer_2021} & 42.46 & 9.89 & 90.46$\pm$0.33 & 83.92$\pm$0.45 \\
DAEFormer \cite{zhang_dual-attention_2024}  & 48.07 & 27.89 & 91.19$\pm$0.31 & 85.40$\pm$0.43 \\
HiFormer \cite{heidari_hiformer_2023} & 25.51 & 8.05 & 91.48$\pm$0.28 & 85.15$\pm$0.40 \\
TransFuse \cite{de_bruijne_transfuse_2021} & 26.25 & 8.82 & 92.03$\pm$0.26 & 86.19$\pm$0.38 \\
D-LKA \cite{azad_beyond_2024}  & 101.64 & 19.92 & 92.83$\pm$0.23 & 87.15$\pm$0.34 \\
SU-Net \cite{li_sunet_2023} & 20.9 & 4.58 & 92.33$\pm$0.24 & 86.58$\pm$0.36 \\
U-Mamba \cite{ma2024u} & 16.40 & 3.51 & 91.77$\pm$0.27 & 86.16$\pm$0.42 \\
SAM2-UNet \cite{xiong2024sam2} & 21.09 & 5.65 & 91.52$\pm$0.29 & 85.88$\pm$0.44 \\
MSCA-Net \cite{sun2023msca} & 27.09 & 12.88 & 91.35$\pm$0.32 & 85.59$\pm$0.48 \\
\midrule
\textbf{ScaleFusionNet (Ours)} & 67.91 & 16.45 & \textbf{92.94$\pm$0.21} & \textbf{87.35$\pm$0.30} \\
\bottomrule
\end{tabular}
\end{table}

\begin{figure}[h]
    \centering
    \includegraphics[width=1.1\textwidth]{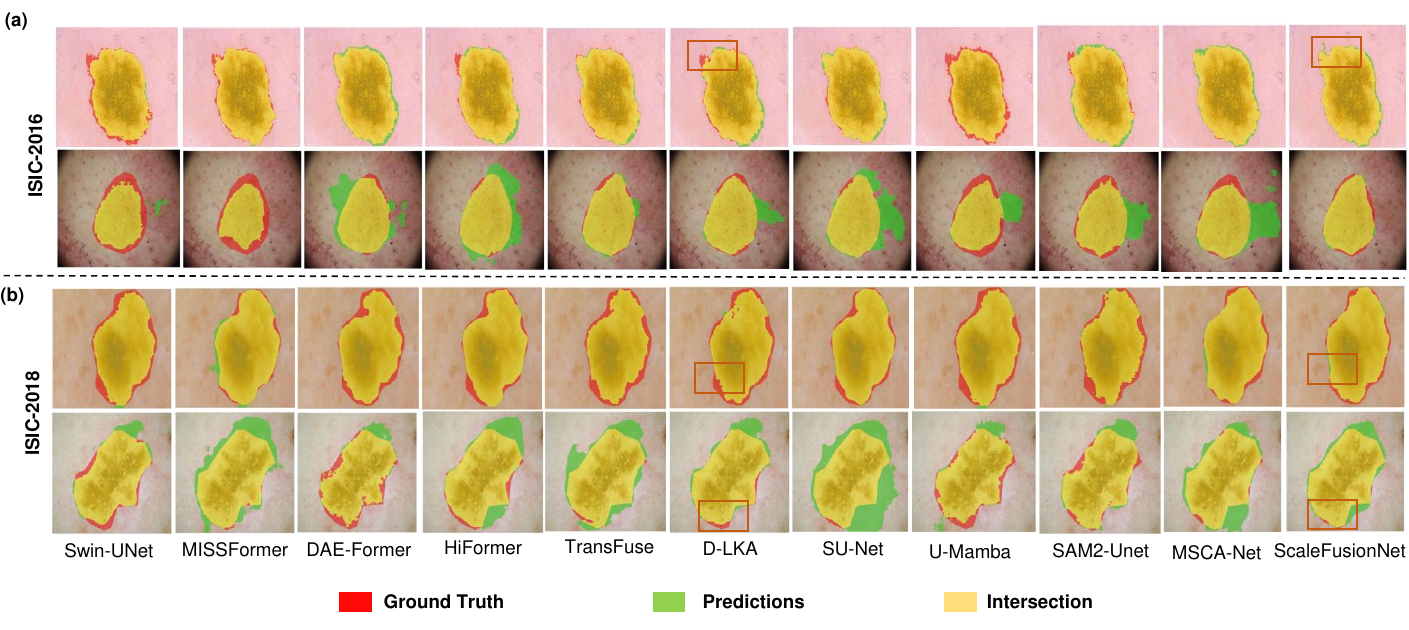}
    \caption{Qualitative comparison of 10 skin lesion segmentation models on ISIC-2016 and ISIC-2018 datasets. Yellow denotes correct prediction, red indicates missed regions (ground truth only), and green highlights false positives(predicted healthy skin as lesion). ScaleFusionNet shows superior overlap with minimal errors. The red square regions illustrate edge refinement capability among the models. }
    \label{fig4}
\end{figure}

To enhance the assessment of model performance, we selected 10 high-performing models for qualitative scrutiny of the experimental outcomes, elucidating the differences among them. The red areas in the illustrations represent ground-truth labels meticulously annotated by experts, reflecting the diagnostic preferences of clinical doctors in real-world scenarios. A larger red area indicates lower model accuracy and poorer discrimination of the affected regions. In contrast, the green areas represent the predicted labels obtained during the model-testing phase. A larger green area suggests that the model has mistakenly segmented healthy skin, which could mislead doctors into treating non-affected areas, especially with destructive procedures, such as lasers or cryotherapy. The yellow areas indicate the overlap between the predicted and ground truth labels; a larger yellow area indicates a more accurate identification of the lesion region by the model. In summary, from the perspective of clinical diagnosis and treatment, smaller green and red areas and a larger yellow area indicate better model performance, enabling more effective segmentation of skin lesions to assist in diagnosis and treatment decisions. The results of the qualitative analysis of the 10 models in the ISIC-2016 dataset are shown in Figure~\ref{fig4}(a). From the first two rows, it is evident that ScaleFusionNet exhibits a broader yellow region than the D-LKA model, indicating that ScaleFusionNet is better at identifying affected areas. Furthermore, ScaleFusionNet's predictions show fewer green and red regions, which reduces the likelihood of misdiagnosis and missed diagnoses in the clinical setting. This difference was even more pronounced in the latter two rows. Although TransFuse had the largest yellow region, it also had the largest green area, indicating a misdiagnosis of healthy regions. Although it covers areas of injury, this misdiagnosis can have serious implications for clinical diagnosis. In comparison with D-LKA, ScaleFusionNet has a similarly sized yellow region but with a smaller misdiagnosis area, aligning better with clinical diagnostic needs.

These results highlight the superior ability of ScaleFusionNet to accurately segment skin lesions while minimizing errors, making it a highly effective tool for clinical applications. Its combination of high accuracy and strong generalization ability positions it as a leading solution for skin lesion segmentation tasks. The model’s ability to preserve fine-grained details and refine the limits of the injury further underscores its potential to improve melanoma diagnosis and treatment in real-world clinical settings.

\subsection{Results on ISIC-2018}

In the experiments conducted on the ISIC-2018 dataset, 14 mainstream medical image segmentation models were selected for comparison with ScaleFusionNet, and more relevant parameters were disclosed. The quantitative analysis results of the ISIC-2018 comparison experiments are listed in Table \ref{tab:performance_comparison}. ScaleFusionNet continued to exhibit competitive results compared to the other 14 medical image segmentation methods, achieving the best results in terms of the DSC metric on ISIC-2018, with most other metrics ranking in the top two. Compared with TransFuse, ScaleFusionNet showed a 0.72\% improvement in the DSC metric while maintaining a consistent IOU level of 0.92\%. Compared with D-LKA, ScaleFusionNet exhibited a 0.06\% improvement in the DSC metric, although it slightly lagged behind D-LKA in the IOU metric. Other metrics, namely, Sensitivity, Specificity, and Accuracy, are also presented in  Table \ref{tab:performance_comparison}, where we can see the performance of ScaleFusionNet compared to other models. Finally,  ScaleFusionNet yielded competitive results with other methods, demonstrating strong performance in skin lesion segmentation.

\begin{table}[htbp]
\centering
\caption{Performance Comparison on ISIC-2018 Dataset. Values are presented as mean $\pm$ standard deviation.}
\label{tab:performance_comparison}
\begin{tabular}{lccccc}
\toprule
\textbf{Methods} & \textbf{DSC} & \textbf{IOU} & \textbf{SE} & \textbf{SP} & \textbf{ACC} \\
\midrule
U-Net \cite{navab_u-net_2015} & $85.45 \pm 0.45$ & $77.33 \pm 0.58$ & $88.00 \pm 0.62$ & $96.97 \pm 0.35$ & $94.04 \pm 0.42$ \\
Att-Unet \cite{oktay_attention_2018} & $85.66 \pm 0.48$ & $77.64 \pm 0.61$ & $86.74 \pm 0.67$ & $98.63 \pm 0.28$ & $93.76 \pm 0.44$ \\
nnU-Net \cite{isensee_nnu-net_2021} & $89.03 \pm 0.38$ & $82.02 \pm 0.49$ & $91.02 \pm 0.54$ & $97.55 \pm 0.32$ & $96.40 \pm 0.37$ \\
Swin-Unet \cite{karlinsky_swin-unet_2023} & $89.31 \pm 0.40$ & $82.14 \pm 0.51$ & $90.99 \pm 0.56$ & $97.20 \pm 0.33$ & $95.99 \pm 0.39$ \\
MISSFormer \cite{huang_missformer_2021} & $89.44 \pm 0.37$ & $82.41 \pm 0.47$ & $90.79 \pm 0.53$ & $96.92 \pm 0.34$ & $96.04 \pm 0.38$ \\
DAEFormer \cite{zhang_dual-attention_2024} & $89.89 \pm 0.35$ & $83.21 \pm 0.44$ & $90.52 \pm 0.50$ & $97.33 \pm 0.31$ & $96.13 \pm 0.36$ \\
HiFormer \cite{heidari_hiformer_2023} & $90.55 \pm 0.33$ & $83.81 \pm 0.42$ & $92.02 \pm 0.48$ & $96.43 \pm 0.32$ & $96.55 \pm 0.34$ \\
TransFuse \cite{de_bruijne_transfuse_2021} & $91.08 \pm 0.31$ & $84.65 \pm 0.39$ & $91.39 \pm 0.46$ & $97.80 \pm 0.29$ & $96.66 \pm 0.32$ \\
D-LKA \cite{azad_beyond_2024} & $91.64 \pm 0.28$ & $85.64 \pm 0.36$ & \textbf{91.94 $\pm$ 0.43} & \textbf{98.20 $\pm$ 0.26} & $96.89 \pm 0.30$ \\
SU-Net \cite{li_sunet_2023} & $90.90 \pm 0.32$ & $84.49 \pm 0.40$ & $90.76 \pm 0.47$ & $97.64 \pm 0.30$ & $96.66 \pm 0.33$ \\
TranSiam \cite{huang_missformer_2021} & $90.44 \pm 0.37$ & $83.45 \pm 0.49$ & $90.83 \pm 0.55$ & $96.92 \pm 0.34$ & $95.04 \pm 0.39$ \\
U-Mamba \cite{ma2024u} & $89.74 \pm 0.36$ & $83.16 \pm 0.45$ & $90.83 \pm 0.52$ & $97.33 \pm 0.31$ & $97.75 \pm 0.35$ \\
SAM2-UNet \cite{xiong2024sam2} & $89.52 \pm 0.38$ & $83.07 \pm 0.46$ & $90.75 \pm 0.54$ & $98.13 \pm 0.28$ & $97.54 \pm 0.37$ \\
MSCA-Net \cite{sun2023msca} & $89.31 \pm 0.39$ & $83.28 \pm 0.45$ & $90.37 \pm 0.55$ & $97.53 \pm 0.31$ & $97.24 \pm 0.38$ \\
\midrule
\textbf{ScaleFusionNet (Ours)} & $\mathbf{91.80 \pm 0.26}$ & $\mathbf{85.57 \pm 0.34}$ & $90.88 \pm 0.41$ & $97.67 \pm 0.25$ & $\mathbf{98.24 \pm 0.28}$ \\
\bottomrule
\end{tabular}
\end{table}

The qualitative analysis results of ISIC-2018 are shown in Figure~\ref{fig4}(b). Visual inspection of the results in the last two rows reveals that, compared with D-LKA, ScaleFusionNet has a larger yellow region and a smaller red region, indicating a higher prediction accuracy, even though it performs slightly worse on the IOU metric. In comparison to TransFuse, while the yellow regions were nearly identical in size, ScaleFusionNet exhibited a smaller range of green areas, demonstrating its superior ability to identify skin lesion regions and reduce the likelihood of misdiagnosis. Similarly, the results in the last two rows further highlight the overall superior performance of the ScaleFusionNet. Based on the experiments using the ISIC-2018 dataset, the D-LKA model and ScaleFusionNet achieved the first and second best performances in terms of the DSC and IOU metrics, respectively. In terms of overall performance, ScaleFusionNet demonstrated better accuracy and boundary fitting. The qualitative analysis of the results, as illustrated in Figure~\ref{fig4}(b), shows that ScaleFusionNet has a smaller green region than D-LKA, indicating a reduced likelihood of misdiagnosis. This further underscores the ability of ScaleFusionNet to accurately segment skin lesions while minimizing errors. These results highlight the superior performance of ScaleFusionNet on the ISIC-2018 dataset, achieving high accuracy in lesion segmentation while maintaining efficiency in terms of parameter count and computational complexity. Its ability to reduce misdiagnosis and improve lesion boundary delineation makes it a highly effective tool for skin lesion analysis, particularly in clinical applications in which precision and efficiency are critical.

\begin{figure}[h]
    \centering
    \includegraphics[width=0.8\textwidth]{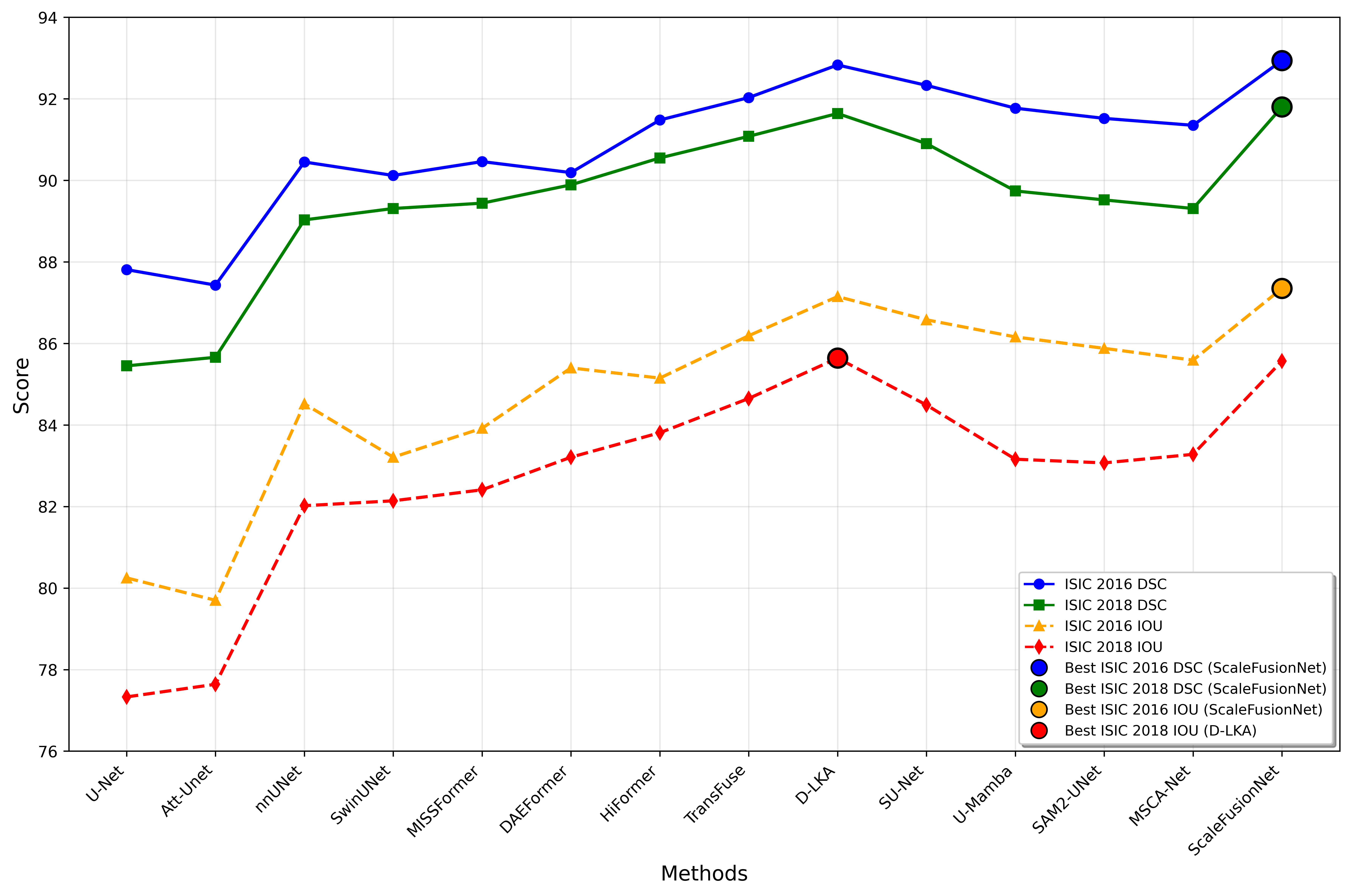}
    \caption{Performance comparison of segmentation models on ISIC-2016 and ISIC-2018 datasets. The DSC (circles) and IOU (squares) scores are shown for various models, with ScaleFusionNet outperforming the other models in both metrics across both datasets.}
    \label{fig6}
\end{figure}

Figure ~\ref{fig6} visually demonstrates the results of various segmentation models in the ISIC 2016 and 2018 datasets, focusing on both DSC and IOU metrics. ScaleFusionNet consistently outperformed the other models, particularly on the ISIC 2016 dataset, where it achieved the highest DSC score. The blue line for ISIC 2016 (DSC) shows a steady increase, with ScaleFusionNet clearly marked as the top performer. Similarly, for ISIC 2018 (DSC), ScaleFusionNet, indicated by the green line, again achieves a leading position, closely followed by models such as D-LKA and TransFuse. The dashed red and yellow lines represent the IOU scores, where ScaleFusionNet excels in the 2016 dataset, further demonstrating its superiority in multiple evaluation metrics. Overall, the graph illustrates the effectiveness of ScaleFusionNet in both DSC and IOU metrics for both datasets, demonstrating it to be a robust model for skin lesion segmentation.

\subsection{External Validation with ISIC-2018 }
\begin{figure}[h]
    \centering
    \includegraphics[width=0.8\textwidth]{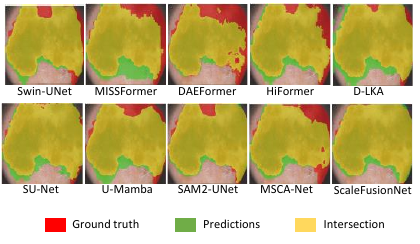}
    \caption{Visual comparison of 10 models on the PH$^2$ dataset using ISIC-2018 pretrained weights. Red shows ground truth, green shows incorrect predictions, and yellow shows accurate segmentation. ScaleFusionNet and SU-Net perform best, with ScaleFusionNet showing better boundary accuracy and fewer false positives.}
    \label{fig7}
\end{figure}

 Independent validation experiments were conducted on the PH$^2$ dataset using the model weights pretrained on ISIC-2018. The results show that ScaleFusionNet demonstrates significant performance improvements compared to the SU-UNet model, with a DSC increase of 0.05\% and an IOU increase of 0.87\%, indicating an improvement over the experiments on the ISIC-2018 dataset. From Table \ref{tab:ph2_results}, it can be observed that ScalefusionNet continues to exhibit competitive results compared to the other 9 medical image segmentation methods, achieving the best results in terms of the DSC metric on the PH$^2$ dataset, with most other metrics ranking in the top two. Based on the external validation experiments using the ISIC2018 dataset, the SUnet model and ScaleFusionNet achieved the first and second best performances in terms of the DSC and IOU metrics, respectively. The qualitative analysis of the external validation experiments is shown in Figure \ref{fig7}. Visual inspection of the results from the external validation experiments using the ISIC-2018 trained weights on the PH$^2$ dataset reveals that the yellow region of the SUnet model was larger, indicating a better ability to accurately predict lesions. In contrast, compared with ScaleFusionNet, the green region in the bottom-left corner is also larger for SUnet, suggesting that ScaleFusionNet performs better in terms of accuracy and fitting to boundaries. Overall, both SUnet and ScaleFusionNet outperformed the other selected comparison models in terms of visualized results based on ISIC-2018 trained weights on the PH$^2$ dataset. This conclusion demonstrates the excellent performance of the ScaleFusionNet skin lesion segmentation method proposed in this paper.

\begin{table}[h]
\begin{tabular}{cccccc}
\toprule
\multicolumn{1}{c|}{\multirow{2}{*}{\textbf{Methods}}} & \multicolumn{5}{c}{\textbf{PH$^2$ (test)}} \\ 
\multicolumn{1}{c|}{} & \textbf{DSC} & \textbf{SE} & \textbf{SP} & \textbf{ACC} & \textbf{IOU} \\
\midrule
Swin-UNet & 90.92$\pm$0.45 & 96.97$\pm$0.32 & 91.42$\pm$0.58 & 93.39$\pm$0.41 & 84.09$\pm$0.67 \\
MISSFormer & 91.9$\pm$0.38 & 97.1$\pm$0.29 & 92.82$\pm$0.45 & 94.24$\pm$0.33 & 85.49$\pm$0.54 \\
DAEFormer & 90.28$\pm$0.52 & 97.41$\pm$0.26 & 90.02$\pm$0.61 & 92.99$\pm$0.48 & 83.37$\pm$0.73 \\
HiFormer & 92.03$\pm$0.41 & 96.6$\pm$0.35 & 93.46$\pm$0.39 & 94.45$\pm$0.37 & 85.88$\pm$0.58 \\
D-LKA & 92.17$\pm$0.36 & 97.3$\pm$0.28 & 93.53$\pm$0.42 & 94.52$\pm$0.34 & 86.14$\pm$0.51 \\
SUnet & 92.32$\pm$0.33 & 98.14$\pm$0.21 & 92.19$\pm$0.47 & \textbf{94.86$\pm$0.31} & 86.23$\pm$0.49 \\
U-Mamba & 92.1$\pm$0.39 & 97.7$\pm$0.25 & 93.34$\pm$0.43 & 94.51$\pm$0.35 & 85.66$\pm$0.55 \\
SAM2-UNet & 91.83$\pm$0.44 & 98.02$\pm$0.22 & 92.46$\pm$0.46 & 94.52$\pm$0.32 & 85.48$\pm$0.62 \\
MSCA-Net & 91.76$\pm$0.47 & 96.97$\pm$0.34 & \textbf{93.79$\pm$0.38} & 94.27$\pm$0.29 & 85.28$\pm$0.66 \\ 
\midrule
\textbf{ScaleFusionNet(Ours)} & \textbf{92.37$\pm$0.31} & \textbf{98.23$\pm$0.19} & 92.44$\pm$0.45 & 94.73$\pm$0.28 & \textbf{87.10$\pm$0.44} \\ 
\bottomrule
\end{tabular}
\caption{Performance comparison on PH$^2$ dataset. Values are presented as mean $\pm$ standard deviation.}
\label{tab:ph2_results}
\end{table}

\subsection{Ablation Study}
To validate the effectiveness of the proposed ScaleFusionNet, a structural ablation study was conducted on the ISIC-2016 dataset, with the DSC used as the primary evaluation metric. The data from the structural ablation studies are listed in Table \ref{tab:msdunet_ablation}. Method 0 represents the segmentation results obtained using only the hybrid architecture, Method 1 represents the experimental results obtained using both the hybrid architecture and the CATM, and Method 2 represents the experimental results obtained using only the hybrid architecture and the AFB.

\begin{figure}[h]
    \centering
    \includegraphics[width=0.8\textwidth]{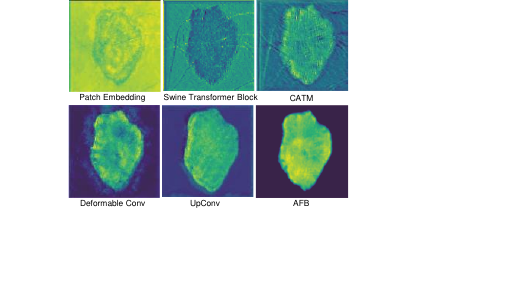}
    \caption{Feature map visualization of different layers in ScaleFusionNet. Each module, including Patch Embedding, Swin Transformer Block, CATM, Deformable Convolution, UpConv, and AFB captures distinct features of the lesion.}
    \label{fig5}
\end{figure}

From Table \ref{tab:msdunet_ablation}, it can be observed that using only the hybrid architecture results in a DSC of 91.76\% on ISIC-2016, which is still superior to most models compared in Table \ref{tab:isic2016}. Method 1, which incorporates the CATM to enhance the skip connection features on top of the hybrid architecture, achieves a DSC of 92.54\% on ISIC-2016. This performance surpasses that of the SU-Net method at 92.33\%, albeit slightly weaker than that of the D-LKA method. Method 2, which introduces only the AdaptiveFusionBlock on top of the hybrid architecture, achieves a DSC performance of 92.62\%. Finally, ScaleFusionNet, which combines the hybrid architecture, CATM, and AdaptiveFusionBlock, achieves a DSC of 92.94\% on ISIC-2016. The excellent performance of ScaleFusionNet is propelled by integrating these three proposed improvement methods. Furthermore, we visualize the features of Stage 0 and the corresponding modules within the same layer to intuitively observe the role of each structure. As shown in Figure~\ref{fig5}, it's clear that the focus on the target area becomes much stronger after it goes through the CATM, following the output of the hybrid architecture. Additionally, each feature map of the four multi-scale branches highlights different attention areas. This observation underscores the emphasis on micro-scale multi-resolution in this study. The micro-scale multi-resolution enables the features passed through the AFB to focus highly on the target area and exhibit excellent fitting to the target boundaries.

\begin{table}[h]
    \centering
    \caption{Structural ablation results of ScaleFusionNet on ISIC-2016. The Swin Transformer Block, CATM, and AFB were tested individually and in combination. }
    \label{tab:msdunet_ablation}
    \begin{tabular}{lccc c}
        \toprule
        Methods & Swin Transformer Block & CATM & AFB & DSC$\uparrow$ \\
        \midrule
        0 & \checkmark & & & 91.76 \\
        1 & \checkmark & \checkmark & & 92.54 \\
        2 & \checkmark & \checkmark & \checkmark & 92.62 \\
        ScaleFusionNet & \checkmark & \checkmark & \checkmark & 92.94 \\
        \bottomrule
    \end{tabular}
\end{table}

These ablation experiments show how important the CATM and AFB are for ScaleFusionNet's performance, and they also highlight the need to improve the encoder's design. The results highlight the model's ability to achieve high accuracy in skin lesion segmentation while maintaining an efficient and balanced architecture.

\section{Conclusion and Future Work}\label{sec5}

This study presents a medical image segmentation model called ScaleFusionNet, which incorporates the CATM and AFB to learn and extract complex features from medical images. Experiments conducted on publicly available datasets demonstrate that ScaleFusionNet achieved competitive results in skin lesion segmentation. These innovative approaches positively impact diagnostic accuracy, guide treatment decisions, and promote further research in the field. However, its computational complexity is higher compared to some methods. This problem mainly comes from using multi-scale feature extraction and cross-attention mechanisms, where the input from different areas differs. One way to solve this is by better feature allocation, like self-selective routing. Also, deformable convolutions might worsen the issue, so simpler methods should be explored in future work. From a clinical perspective, a reliable medical image segmentation method must not only provide high-quality segmentation results but also deliver corresponding uncertainty metrics. ScaleFusionNet is an important advancement in skin lesion segmentation, as it uses adaptive multi-scale fusion and cross-attention mechanism to achieve accurate and strong results. However, future work should address computational efficiency and incorporate uncertainty quantification to further enhance its clinical applicability and reliability.

\backmatter

\bmhead{Acknowledgements}

This work was supported in part by the UK EPSRC under Grant EP/P009727/2, and the Leverhulme Trust under Grant RF-2019-492. The authors also express their appreciation to Taif University, Saudi Arabia, for supporting this work through project number (TU-DSPP-2024-229).

\section*{Declarations}


 \textbf{Availability of Data and Materials:} The datasets analyzed in the current study are available in the 2016 and 2018 repositories at the following link: \href{https://challenge.isic-archive.com/data/}{https://challenge.isic-archive.com/data/}.

\noindent\textbf{Conflict of interest:}  The authors declare no competing interests.

\noindent\textbf{Source Code:}  The implementation code can be found by clicking on this \href{https://github.com/sqbqamar/ScaleFusionNet}{link}.

\noindent\textbf{Ethical approval:} This study did not involve animals or human participants.

\section*{Author Contribution}
SQ conceived the study. SQ designed, developed, and evaluated the algorithm's performance. SFQ, RA, GMA, and RJ validated the algorithm. All authors contributed to writing the manuscript and approved the final version.



\bibliography{skin}

\end{document}